\documentclass[prl,twocolumn,a4paper,showpacs,superscriptaddress]{revtex4}
\usepackage{amsmath}
\usepackage{amsfonts}
\usepackage{graphicx}
\usepackage{longtable}
\newcommand{\be}{\begin{equation}}
\newcommand{\ee}{\end{equation}}

\newcommand{\ba}[1]{\left(\begin{array}{#1}}
\newcommand{\ea}{\end{array}\right)}
\begin{document}

\title{Interconvertibility and irreducibility  of  permutation symmetric three qubit pure states} 

\author{A. R. Usha Devi}
\email{arutth@rediffmail.com}
\affiliation{Department of Physics, Bangalore University, 
Bangalore-560 056, India}
\affiliation{Inspire Institute Inc., McLean, VA 22101, USA.}
\author{Sudha}
\affiliation{Department of Physics, Kuvempu University, 
Shankaraghatta, Shimoga-577 451, India}
\author{A. K. Rajagopal} 
\affiliation{Inspire Institute Inc., McLean, VA 22101, USA.}
\date{\today}

\begin{abstract} 
A novel use of Majorana geometric representation brings out distinct entanglement families 
of permutation symmetric states of qubits.  The paradigmatic W and  GHZ (Greenberger-Horne-Zeilinger)   states 
of three qubits respectively contain two and three independent Majorana spinors. 
Another unique state with three distinct Majorana spinors --  constructed through a permutation symmetric 
superposition of  two {\em up} qubits and one {\em down} qubit  (W state) and its obverse state 
($\bar {\rm W}$) exhibits genuine three-party entanglement, which is robust under loss of a qubit. 
While the GHZ state has irreducible correlations and cannot be determined from its parts, we show here that the 
correlation information of the W-superposition state is imprinted uniquely in its two party reduced states. 
This striking example sheds light on the contrasting irreducibility features of interconvertible states. 
\end{abstract}
\maketitle

Understanding different kinds of correlations exhibited by multiparticle quantum systems is one of the 
central issues of importance in quantum information science~\cite{Niel}. Two $N$-party {\em pure}  
states $\vert\phi\rangle$, $\vert \psi\rangle$  are interconvertible, with  non-zero probability of success, 
by means of stochastic local operations and classical communications (SLOCC)  if and only if there exists an 
invertible local operation (ILO)~\cite{Dur} $A_1\otimes A_2\otimes \ldots \otimes A_N$ such that $\vert 
\phi\rangle=A_1\otimes A_2\otimes \ldots \otimes A_N\vert\psi\rangle.$ Local unitary (LU) operations $U_1\otimes 
U_2\otimes \ldots U_N$ form a subset of ILOs and  two multiparty states related to each other through LU possess 
exactly {\em same} amount of entanglement, while interconvertibility via ILO correspond, in general, to modified 
entanglement features. General classification of distinct kinds of entanglement that are 
{\em inequivalent} under SLOCC has gained increasing attention in recent years~\cite{Dur,Ver,Lamata, Kraus}.    
It is found to be convenient to address this issue by restricting to special classes of states exhibiting some 
particular symmetry, in order to tackle the algebraic complexity associated with exponentially increasing size 
of the Hilbert space. In the illuminating case of  $N$-qubits obeying permutation symmetry,  it is sufficient 
to search for identical ILOs of the form $A\otimes A\otimes\ldots \otimes A$ to verify the SLOCC equivalence of 
two pure states~\cite{Solano,Bastin}.  The significance of such considerations is catiching up and  innovative 
experimental schemes to generate a large variety of multiqubit states has been proposed 
recently~\cite{newexptl}. The representation proposed by Majorana~\cite{Majorana} as early as 1932 offers a 
deeper understanding on how different entanglement families emerge, depending on the number and arrangement of 
the independent spinors (qubits) constituting the {\em pure} symmetric state ~\cite{Solano}. 

``Can higher order correlations follow entirely from lower order ones?" is also a question 
of fundamental interest both from the modern perspective of quantum information science~\cite{SP} as well as in 
many body physics~\cite{Coleman}. Identifying the extent to  which correlation content of a $n$-party state 
can/cannot be ascribed to that within groups of fewer than $n$ parties and finding methods of characterizing 
{\em irreducibility} of $n$-party correlations would indeed shed light on different possible types of 
correlations that quantum states can exhibit. Linden, Popescu and Wootters~\cite{SP} proved an interesting 
result that the correlation content of {\em almost all} pure states, shared by $n$ parties, is already contained 
in their reduced states i.e., the  correlations within a generic $n$-party pure state are {\em 
reducible}~\cite{SP}.  

 ``Do interconvertible states possess similar {\em irreducibility} features?" is a related important question 
 not addressed before. In the present Letter we draw attention to distinct entanglement families of 
 permutation symmetric three qubit symmetric and explicitly show that  two different states belonging 
 to the same SLOCC class can exhibit contrasting features as 
far as irreducibility~\cite{SP} of their correlations is concerned. 
A general discussion on how/if the $N$-party correlations within a pure symmetric $N$-qubit state --  
belonging to a specific entanglement family  --  follow  from lower order correlations carried by the 
reduced subsystems, will be presented elsewhere~\cite{ARUS}.

\noindent {\em Majorana representation:} A system of $N$-qubits obeying exchange symmetry gets restricted to 
a $(N+1)$ dimensional Hilbert space spanned by the basis vectors $\{\vert N/2,k- N/2\rangle, k=0,1,2,
\ldots N \}$ where,    
\begin{eqnarray}
\label{Dicke} 
\vert N/2, k-N/2\rangle &=&\frac{1}{\sqrt{^N C_k}}\,[\vert \underbrace{0_1, 0_2, \ldots}_{k\ {\rm times}}\, ,  
\underbrace{1_1, 1_2, \ldots}_{N-k\ {\rm times}}\rangle \nonumber \\ 
&&  \ \ \ \ \ \ \ \ \ \ +\ {\rm Permutations}\ ] 
\end{eqnarray}
are the $N+1$ Dicke  states -- expressed in the standard qubit basis $\vert 0\rangle,\ \vert 1\rangle$. (Here, 
 $^N C_k=\frac{N!}{k!\,(N-k)!}$ denotes the binomial coefficient). 
 
An arbitrary pure symmetric state,    
\begin{equation}
\label{sympure1} 
\vert \Psi_{\rm sym}\rangle =\sum_{k=0}^{N}\, d_k\, \vert N/2, k-N/2\rangle,
\end{equation}
is specified by the $(N+1)$ complex coefficients $d_k.$ Eliminating an overall phase and normalizing  the 
state (i.e., $\sum_{k=0}^{N}\, \vert d_k\vert^2=1$) implies that $N$ complex parameters are required to 
completely characterize a pure symmetric state of $N$ qubits.  

Alternately, Majorana~\cite{Majorana} expressed the  pure state $\vert \Psi_{\rm sym}\rangle$ as a superposition 
of  {\em symmetrized} states of  $N$ spin-$1/2$ particles: 
\begin{equation}
\label{Maj}
\vert \Psi_{\rm sym}\rangle={\cal N}\, \sum_{P}\, \hat{P}\, \{\vert \epsilon_1, \epsilon_2, 
\ldots  \epsilon_N \rangle\} 
\end{equation} 
where $\vert \epsilon_s\rangle=\cos(\beta_s/2)\, e^{-i\alpha_s/2}\, \vert 0_s\rangle
+\sin(\beta_s/2)\, e^{i\alpha_s/2}\, \vert 1_s\rangle$,  $s=1,2,\ldots, N$ denote spinors 
constituting the state $\vert \Psi_{\rm sym}\rangle$; $\hat{P}$ denotes the set of all $N!$ 
permutations and ${\cal N}$ corresponds to an overall normalization factor. So,  $N$ complex parameters 
$z_s=\tan\frac{\beta_s}{2}e^{i\alpha_s}$ offer an alternate parametrization for the pure $N$ qubit 
symmetric state. As an identical rotation $R\otimes R\ldots \otimes R$ on the symmetric state $\vert \Psi_{\rm 
sym}\rangle$ transforms it into another symmetric state, choosing $R=R_l^{-1}\equiv R^{-1}(\alpha_l,\beta_l,0)$ 
(where $(\alpha_l,\beta_l,0)$ denote the Euler angles of rotation~\cite{Rose}) such that it aligns one of the 
constituent spinors say, $\vert \epsilon_l\rangle$, along the positive $z$-direction i.e., 
$R_l^{-1}\vert\epsilon_l\rangle=\vert 0_l\rangle$, results in the following identification~\cite{Majnote}, 
\begin{equation}
\label{rl}
\langle 1_1,1_2,\ldots 1_N\vert R_l^{-1}\otimes R_l^{-1}\ldots \otimes R_l^{-1}\, \vert \Psi_{\rm 
sym}\rangle\equiv 0
\end{equation}
Eq.~(\ref{rl}) holds good for any identical rotations $R^{-1}_s\otimes 
R^{-1}_s\otimes\ldots\otimes 
R^{-1}_s,\ \ s=1,2,\ldots, N,$ orienting {\em any} one of the constituent  
qubits in the positive $z$-direction. In other words, there exist $N$ rotations 
$R^{-1}_s=R^{-1}(\alpha_s,\beta_s,0), s=1,2,\ldots , N$,
 which lead to the same result as in (\ref{rl}). 
 
In terms of the alternate representation (\ref{sympure1}), we obtain 
\begin{eqnarray}
\label{cmj0}
\sum_{k=0}^{N}\, d_k\,\langle N/2,-N/2\vert {\cal R}^{-1}  \vert N/2,k-N/2\rangle &&\nonumber \\ 
\hskip 0.2in =  \sum_{k=0}^N\, d_k\  D^{N/2*}_{k-N/2,-N/2}(\alpha_s,\beta_s,0)&=&0,
\end{eqnarray}
where 
$D^{N/2*}_{k-N/2,-N/2}(\alpha_s,\beta_s,0)=\langle N/2,-N/2\vert{\cal R}^{-1}_s (\alpha_s,\beta_s,0)
 \vert N/2,k-N/2\rangle,$ (where  ${\cal R}^{-1}(\alpha_s,\beta_s,0)~=~R_s^{-1}\otimes 
R_s^{-1}\ldots \otimes R_s^{-1}$ represents the collective rotation 
in the $N+1$ dimensional symmetric subspace) denotes the elements of the rotation matrix 
 in the Wigner-$D$ representation~\cite{Rose}. 
Substituting the explicit form of the $D$-matrix~\cite{Rose},   
$D^{N/2*}_{k-N/2,-N/2}(\alpha_s,\beta_s,0)=(-1)^k, \sqrt{^NC_{k}}\, 
\left(\cos\frac{\beta_s}{2}\right)^{(N-k)}\, \left(\sin\frac{\beta_s}{2}\right)^k \, e^{i(k-N/2)\alpha_s}$,
in (\ref{cmj0}) and simplifying, we obtain
\begin{eqnarray}
\label{cmj02}
{\cal A}\, \sum_{k=0}^N (-1)^k\,\sqrt{^N\, C_k}\,  d_k\,  \, z^{k}&=&0\,  
\end{eqnarray} 
where $z=\tan\frac{\beta_s}{2}\,e^{i\, \alpha_s},$ and 
${\cal A}=\cos^N\frac{\beta_s}{2}\,e^{-i\alpha_s\, N/2}$. In other words, given the parameters $d_k$, 
the $N$ roots $z_s, s=1,2,\ldots N$ of the Majorana polynomial 
$P(z)=\sum_{k=0}^N\, (-1)^k\, \sqrt{^N\, C_k}\,  d_k\,  \, z^{k}$
determine the orientations $(\alpha_s,\beta_s)$ of the  spinors 
constituting the  $N$-qubit symmetric state.  
The list of degneracy numbers $\left\{n_1,n_2,\ldots n_d;  \sum_i 
n_i=N\right\}$ -- where $d$ denotes the number of distinct spinors $\vert \epsilon_1\rangle, \vert 
\epsilon_2\rangle,\ldots, \vert \epsilon_d\rangle,$
(which are determined by the independent solutions of the Majorana polynomial) respectively appearing   $n_1\geq 
n_2\geq \ldots \geq n_d$ times,   is employed by Bastin et. al.~\cite{Solano} to classify  pure symmetric states 
into 
different families denoted by $\{D_{n_1,n_2,\ldots n_d}\}$. For e.g.,    
when all the $N$ solutions of the Majorana polynomial are identically equal, the symmetric state is given by 
$\vert \Psi_{\rm sym}^{(N)}\rangle=\vert \epsilon,\epsilon,\ldots \epsilon\rangle$, with 
the degeneracy $N$; the symmetric state  is then said to belong to the family of 
separable  states denoted by $\{D_N\}$.  The symmetric states with two distinct spinors have the form, 
$\vert \Psi_{\rm sym}^{(n_1,n_2)}\rangle={\cal N}\,[\vert \underbrace{\epsilon_1, \epsilon_1,\ldots 
\epsilon_1}_{n_1}\, , \underbrace{\epsilon_2, \epsilon_2,\ldots \epsilon_2}_{n_2}\rangle+{\rm \, 
Permutations\,}]$; 
$n_1\geq n_2, n_1+n_2=N$, and they belong to the  family $\{D_{n_1,n_2} \}$ etc. Clearly, two pure symmetric 
states belonging to different classes can not be related to one another by an identical ILO $A\otimes 
A\otimes\ldots \otimes A$ and hence, they are inequivalent under SLOCC~\cite{Solano}. 

\noindent {\em Irreducibility of pure three qubit symmetric states  -- (i)~the SLOCC 
class $\{D_{2,1}\}$}: 

Let us consider a three qubit pure symmetric state with two of the spinors $\vert \epsilon_1\rangle,\vert 
\epsilon_2\rangle,$ distinct i.e., the states belonging 
to the SLOCC class $\{D_{2,1}\}.$  The symmetrized three qubit state (see Eq.~(\ref{Maj})) is given by,   
\begin{eqnarray}
\label{3qubitW}
\vert D_{2,1}\rangle&=&{\cal N}\, \sum_{P}\hat{P}\left\{\vert \epsilon_1, \epsilon_1, 
\epsilon_2\rangle\right\}\nonumber \\
&=&{\cal N}\,\left[ \vert \epsilon_1, \epsilon_1, \epsilon_2\rangle 
 +\vert \epsilon_2, \epsilon_1, \epsilon_1\rangle + \vert \epsilon_1, \epsilon_2, \epsilon_1\rangle \right],  
\end{eqnarray}
 where the spinors 
$\vert\epsilon_s\rangle= R_s\vert 0_s\rangle=\cos(\beta_s/2)\, e^{-i\alpha_s/2}\, \vert 0_s\rangle
+\sin(\beta_s/2)\, e^{i\alpha_s/2}\, \vert 1_s\rangle,\ s=1,2.$  
We simplify (\ref{3qubitW}) as follows: 
\begin{eqnarray}
\label{3W2}
\vert D_{2,1}\rangle=
{\cal N}\,\left[\sum_P {\hat P}\,\{R_1\otimes R_1\otimes R_2\}\right] \vert 
0_1,0_2,0_3\rangle\hskip 0.5in \nonumber \\ 
\ \ \ = {\cal N}\, R_1\otimes R_1\otimes R_1\, \sum_P {\hat P}\,\{I\otimes I\otimes R_1^{-1}R_2\}  \vert 
0_1,0_2,0_3\rangle. \nonumber  
\end{eqnarray} 
Denoting $R^{-1}_1\otimes R^{-1}_1\otimes R^{-1}_1\, \vert D_{2,1}\rangle=\vert D'_{2,1}\rangle,$ where 
$\vert D'_{2,1}\rangle$  is local unitarily related to  $\vert D_{2,1}\rangle$ and belongs to the same SLOCC 
class  $\{D_{2,1}\}$, we get, 
\begin{eqnarray}
\vert D'_{2,1}\rangle&=&{\cal N}\,\left[ \vert 0_1,0_2,\epsilon'_2\rangle 
 +\vert \epsilon'_2, 0_2, 0_3\rangle + \vert 0_1,  \epsilon'_2, 0_3\rangle \right]     
\end{eqnarray} 
with $\vert\epsilon'_2\rangle=R_1^{-1}\,R_2\vert \epsilon_2\rangle=c_0\,\vert 0\rangle +\, c_1\,\vert 1\rangle$,  
$c_0^2+c_1^2=1.$ We thus obtain the states belonging to the class of 3-qubit symmetric pure states  $\vert 
D_{2,1}\rangle$, with only two distinct Majorana spinors (upto an identical local unitary transformation) 
as~\cite{WDicke}, 
\begin{equation}
\label{Wfin}
\vert D'_{2,1}\rangle= a\,  \vert 0_1,0_2,0_3\rangle + \sqrt{3}\, b\, \vert {\rm W}\rangle
\end{equation}
where $a=\frac{\sqrt{3}\, c_0}{\sqrt{3\vert c_0\vert^2+\vert c_1\vert^2}},$ 
$b=\frac{ c_1}{\sqrt{9\,\vert c_0\vert^2+3\,\vert c_1\vert^2}},$ and 
\begin{equation} 
\label{W}
\vert {\rm W}\rangle=\frac{1}{\sqrt{3}}[\vert 1_1,0_2,0_3\rangle +\vert 0_1,1_2,0_3\rangle+\vert 
0_1,0_2,1_3\rangle
\end{equation}
is the 3-qubit W state.  
We now show that 3-party correlation in the state (\ref{Wfin}) is {\em reducible} as the state is {\em uniquely} determined in terms of its 2-party reduced density matrices.    

Let us suppose  that a mixed  three qubit state $\omega$ too shares the  two-qubit reduced system $\rho_{12}$, 
same as that of $\vert D'_{2,1}\rangle$ i.e., $\rho_{12}={\rm Tr}_{3}[\vert D_{2,1}'\rangle\langle 
D_{2,1}'\vert]={\rm Tr}_{3}\omega.$  The mixed state $\omega$ may always be thought of as a reduced system of a 
pure state  $\vert\Omega\rangle$ of the three qubits and an environment $E$ such that ${\rm 
Tr}_{E}[\vert\Omega\rangle\langle \Omega\vert]=\omega$ and so, the two party reduced state $\rho_{12}$ can be 
expressed  as  $\rho_{12}={\rm Tr}_{3}[\vert \Omega\rangle\langle \Omega\vert].$
Consistency between these two lead to the desired result.  
The two party reduced  state $\rho_{12}$ of (\ref{Wfin}) is a rank-2 mixed state 
\begin{eqnarray}
\label{wp2}
\rho_{12}&=&{\rm Tr}_3[\vert D'_{2,1}\rangle\langle D'_{2,1}\vert]=\vert \phi_0\rangle\langle \phi_0\vert + 
\vert \phi_1\rangle\langle \phi_1\vert, 
\end{eqnarray}
where
$\vert \phi_0\rangle= a\, \vert 0_1,0_2\rangle + b\, (\vert 1_1, 0_2\rangle
 +\vert 0_1,1_2\rangle),\  
   \vert \phi_1\rangle= b\, \vert 0_1,0_2\rangle.$  
And in order that the  pure state $\vert\Omega\rangle$ (or the mixed state $\omega$) too shares the same two 
qubit reduced state $\rho_{12}$, we should have 
\begin{eqnarray}
\label{wp3}
\vert\Omega\rangle&=&\vert \phi_0\rangle\vert E_0\rangle +\vert \phi_1\rangle\vert E_1\rangle, \ \  
\langle E_i\vert E_j\rangle=\delta_{i,j},\ \ 
\end{eqnarray}     
The states $\vert E_{0}\rangle, \vert E_{1}\rangle$ are the ones containing the third qubit and the rest of the 
environment:    
\begin{eqnarray*}   
\vert E_0\rangle=\vert 0_3\rangle\,  \vert e_{00}\rangle+\vert 1_3\rangle\,  \vert e_{01}\rangle,\ 
 \vert E_1\rangle=\vert 0_3\rangle\,  \vert e_{10}\rangle+\vert 1_3\rangle\,  \vert e_{11}\rangle.   
\end{eqnarray*}
Thus, the state $\vert \Omega\rangle$ can be expressed explicitly as, 
\begin{eqnarray}
\label{omega}
\vert \Omega\rangle&=& \vert 0_1,0_2,0_3\rangle\, [a\,\vert e_{00}\rangle+b\, \vert e_{10}\rangle] 
+ \vert 
0_1,0_2,1_3\rangle\, [a\vert e_{01}\rangle\nonumber \\
&&+b\vert e_{11}\rangle]+ b\,  \vert 0_1,1_2,0_3\rangle\,\vert e_{00}\rangle  
 \ +b\,  \vert 0_1,1_2,1_3\rangle\,\vert e_{01}\rangle\nonumber \\ 
 &&+ b\,  \vert 1_1,0_2,0_3\rangle\,\vert e_{00}\rangle  
+ b\,  \vert 1_1,0_2,1_3\rangle\,\vert e_{01}\rangle
\end{eqnarray} 
Demanding that the reduced system $\rho_{13}$ of $\vert D_{2,1}'\rangle$  too is shared by $\vert 
\Omega\rangle$ imposes further restrictions~\cite{noteAKR}. 

We first consider the matrix element $\langle 1_1,1_3\vert\rho_{13}\vert 1_1,1_3\rangle$, evaluated from both  
$\vert\Omega\rangle$ (see (\ref{omega})) and $\vert D_{2,1}'\rangle$  (see (\ref{Wfin})): 
We get
 $\langle 1_1,1_3\vert{\rm Tr}_{2,E}\, [\vert\Omega\rangle\langle\Omega\vert]\, \vert 1_1,1_3\rangle=\vert 
b\vert ^2\, \langle e_{01}\vert e_{01}\rangle.$  
But  from (\ref{Wfin}) we obtain  the value zero for this matrix element.  
Hence, $\vert e_{01}\rangle\equiv 0.$ 
Futher, from the orthonormality relations (\ref{wp3}) 
 we obtain $\langle e_{00}\vert e_{00}\rangle=1,$ 
 and $\langle e_{00}\vert e_{10}\rangle=0.$ 

Similarly, we get $\langle 0_1,1_3\vert{\rm Tr}_{2,E}\, [\vert\Omega\rangle\langle\Omega\vert]\, \vert 
1_1,0_3\rangle=\vert b\vert^2\, \langle e_{11}\vert e_{00}\rangle$,  
whereas  (\ref{Wfin}) leads to the value $\vert b\vert^2$, implying that $\vert e_{11}\rangle\equiv \vert 
e_{00}\rangle.$ 
 
Finally,   
$\langle 0_1,0_3\vert{\rm Tr}_{2,E}\, [\vert\Omega\rangle\langle\Omega\vert]\, \vert 0_1,0_3\rangle=\vert 
b\vert^2\, \langle e_{10}\vert e_{10}\rangle+ \ \vert a\vert^2 +\vert b\vert^2$  
is  compared it with its value $\vert a\vert^2+\vert b\vert^2,$  found  from (\ref{Wfin}) to infer that  $\vert 
e_{10}\rangle=0$. 
Thus, we deduce
$\vert\Omega\rangle=\vert D'_{2,1}\rangle\, \vert e_{00}\rangle.$
In other words, the {\em only} extended state  that is consistent 
with the two party reduced states (of all the qubit pairs $(1,2), (2,3)$, 
and $(1,3)$)  of the pure three qubit state $\vert D'_{2,1}\rangle$ is the direct product state  
$\vert\Omega\rangle=\vert D'_{2,1}\rangle\, \vert e_{00}\rangle$,  establishing  the uniqueness of the ``whole 
with its parts" in the case of three qubit states belonging to the SLOCC class $\{D_{2,1}\}$. (Note that we have  
employed only two  of the reduced states $\rho_{12}, \rho_{13}$ to reach this conclusion).         

\noindent{\em (ii) States belonging to the SLOCC class $\{D_{1,1,1}\}$:} Are the three qubit states consisting 
of all three distinct Majorana spinors determined by its reduced systems uniquely? To address this question, we 
consider two specific examples of the SLOCC family $\{D_{1,1,1}\}$,  the first being the three qubit 
GHZ state,  
\begin{equation}
\label{GHZ}
\vert {\rm GHZ}\rangle= \frac{1}{\sqrt{2}}\, [\vert 0_1,0_2,0_3\rangle +  \vert 1_1,1_2,1_3\rangle]. 
\end{equation}      
The Majorana polynomial equation (\ref{cmj02}) for this state has a simple structure $1-z^3=0$, solutions of 
which are cube roots of unity $\omega, \omega^2,\omega^3=1$ and the corresponding spinors constituting the state 
are readily identified to be 
$\vert \epsilon_1\rangle =\frac{1}{\sqrt{2}}[\vert 0\rangle+\omega\, \vert 1\rangle]$,
$\vert \epsilon_2\rangle =\frac{1}{\sqrt{2}}[\vert 0\rangle+\omega^2\, \vert 1\rangle]$,
$\vert \epsilon_3\rangle =\frac{1}{\sqrt{2}}[\vert 0\rangle+ \vert 1\rangle].$ 
 GHZ state is fragile under the loss of a qubit, with vanishing pairwise concurrence~\cite{Wot} for any pairs of 
two qubit reduced density matrices; but it exhibits genuine three-party entanglement~\cite{Dur,RaRe} with the 
maximum tangle~\cite{Kun}  $\tau=1$.  The state exhibits irreducible three party correlations which can not be  
determined by its reduced states~\cite{SP,Walck}.

\noindent {\em   A special state of three qubits}: We now consider the state   
\begin{equation}
\label{Wsup}
\vert \eta\rangle = \frac{1}{\sqrt{2}} [\vert {\rm W}\rangle +  \vert{\rm \bar{W}}\rangle]. 
\end{equation}         
This  is a superposition of the three qubit W state (\ref{W})  
 and its obverse state 
$\vert{\rm \bar{W}}\rangle=\frac{1}{\sqrt{3}}[\vert 1_1,1_2,0_3\rangle+\vert 1_1,0_2,1_3\rangle+\vert 
0_1,1_2,1_3\rangle].$ 
The state $\vert \eta\rangle$ has genuine three party entanglement, quantified in terms of the tangle 
$\tau=1/3$, and it is also robust under the loss of qubits -- as reflected through the concurrence $C=1/3$ for 
any pairs of two qubits.  The three qubit symmetric state $\vert \eta\rangle$ given by (\ref{Wsup}) satisfies the Majorana polynomial 
equation $z(z-1)=0$  and the corresponding spinors constituting the state are 
$\vert \epsilon'_1\rangle =\vert 1\rangle,$\ 
$\vert \epsilon'_2\rangle =\frac{1}{\sqrt{2}}[\vert 0\rangle+ \vert 1\rangle],$ 
$\vert \epsilon'_3\rangle = \vert 0\rangle.$ 
  
\noindent {\em Contrasting $\vert {\rm GHZ}\rangle$ and $\vert \eta \rangle$}: 
While the entanglement features  of the states $\vert {\rm GHZ}\rangle$ and the W superposition 
state $\vert \eta\rangle$ appear to be different, both the states  belong to the same SLOCC class and 
one can be {\em locally} converted from another,  with some finite probability, 
 so that $\vert {\rm GHZ}\rangle=A\otimes A\otimes A\, \vert \eta\rangle$,  where 
$A=\left(\begin{array}{cc} 1 & \omega \\ 1 & \omega^2\, \end{array}\right).$
The corresponding Majorana spinors  
of the states $\vert \eta\rangle$ and $\vert{\rm GHZ}\rangle$ are  related to each other 
up to an overall factor:   
$A\, \vert\epsilon_1'\rangle=\sqrt{2}\omega\, \vert\epsilon_1\rangle,$   
$A\, \vert\epsilon_2'\rangle=-\omega^2\, \vert\epsilon_2\rangle,$ 
and $A\, \vert\epsilon_3'\rangle=   \sqrt{2}\,  \vert\epsilon_3\rangle$.

Eventhough the  GHZ and  W superposition states are candidates of the same entanglement family, they do exhibit 
 contrasting irreducibility features in that the state $\vert\eta\rangle$ gets entirely determined by its parts 
-- whereas the reduced two party states of the GHZ state are separable and so, the three party correlation 
information is {\em not}  imprinted in them.  We proceed to show explicitly that the higher order correlation in 
the W superposition state $\vert \eta\rangle$ is captured uniquely by its two qubit reduced states.   

Following the procedure outlined for the states of the SLOCC class $\{D_{2,1}\},$ we suppose that 
a mixed  three qubit state $\gamma$ too has the same two-qubit reduced system $\varrho_{12}$, 
as that of $\vert \eta\rangle.$ Denoting the pure state $\vert \Gamma\rangle$ to be  
containing the three qubits and the environment such that  ${\rm Tr}_{E}[\vert\Gamma\rangle\langle 
\Gamma\vert]=\gamma$, the two party reduced state $\varrho_{12}$ can be expressed as  $\varrho_{12}={\rm 
Tr}_{3, E}[\vert \Gamma\rangle\langle \Gamma\vert].$ The two qubit reduced system $\varrho_{12}$ of the pure  
state $\vert\eta\rangle$ is a rank-2 state given by,  
\begin{eqnarray}
\label{ws2}
\varrho_{12}&=&\vert \chi_0\rangle\langle \chi_0\vert + 
\vert \chi_1\rangle\langle \chi_1\vert, 
\end{eqnarray}
where
$\vert \chi_0\rangle=\frac{1}{\sqrt{6}}[|1_1,0_2\rangle+|0_1,1_2\rangle+|1_1,1_2\rangle],$ and  
$   \vert \chi_1\rangle=\frac{1}{\sqrt{6}}[|0_1,0_2\rangle+|0_1,1_2\rangle+|1_1,0_2\rangle]$. 
Given that the two party reduced state $\varrho_{12}$ also belongs to the extended pure state 
$\vert\Gamma\rangle$ (or of the mixed state $\gamma$) of the three qubits and the environment, we must have  
\begin{eqnarray}
\label{ws3}
\vert\Gamma\rangle&=&\vert \chi_0\rangle\vert E_0\rangle +\vert \chi_1\rangle\vert E_1\rangle, \ \  
\langle E_i\vert E_j\rangle=\delta_{i,j},\ \ 
 \end{eqnarray}     
In terms of the basis states of qubit 3, the states of the environment 
$\vert E_{0}\rangle, \vert E_{1}\rangle$ are given by
$\vert E_0\rangle=\vert 0_3\rangle\,  \vert e_{00}\rangle+\vert 1_3\rangle\,  \vert e_{01}\rangle$ and  
$\vert E_1\rangle=\vert 0_3\rangle\,  \vert e_{10}\rangle+\vert 1_3\rangle\,  \vert e_{11}\rangle$.   
Now, demanding that the reduced system $\varrho_{13}$ of $\vert \eta\rangle$ is also shared by  $\vert 
\Gamma\rangle$ leads to further constraints. 

First we compare  $\langle 0_1,1_3\vert \varrho_{13}\vert 0_1,1_3\rangle$,  from the states 
(\ref{Wsup}) and (\ref{ws3}): We have,  
$\langle 0_1,1_3|{\rm Tr}_{2}\,[\vert\eta\rangle \langle\eta\vert]\, |0_1,1_3\rangle= 
\frac{1}{3}$ and 
$\langle 0_1,1_3|{\rm Tr}_{2,E}\,[\vert\Gamma\rangle \langle\Gamma\vert]\, |0_1,1_3\rangle= 
\frac{1}{6}\langle e_{01}|e_{01}\rangle+\frac{1}{3}\langle e_{11}|e_{11}\rangle$ 
leading to  
$\langle e_{01}|e_{01}\rangle+2\langle e_{11}|e_{11}\rangle=2.$

Next, we compare  $\langle 1_1,1_3|\varrho_{13}|1_1,1_3\rangle$ evaluated from the states 
$\vert\eta\rangle$ and $\vert\Gamma\rangle$:  
We get, 
$\langle 1_1,1_3|{\rm Tr}_{2, E}\, [|\Gamma\rangle \langle\Gamma|]\,\vert 1_1,1_3\rangle=
\frac{1}{3}\langle e_{01}|e_{01}\rangle+\frac{1}{6}\langle e_{11}|e_{11}\rangle$ 
and $\langle 1_1,1_3|{\rm Tr}_{2}\, [|\eta\rangle \langle\eta|\vert\, |1_1,1_3\rangle=\frac{1}{6}$ 
implying,  
$2\langle e_{01}|e_{01}\rangle+\langle e_{11}|e_{11}\rangle=1.$ 
From these relations we obtain $\langle e_{11}|e_{11}\rangle=1,$ \ 
$\langle e_{01}|e_{01}\rangle=0$ (or $\vert e_{01}\rangle\equiv 0$).
Further, from the orthonormality (\ref{ws3}) 
it follows that $\langle e_{00}\vert e_{00}\rangle=1,$ and  
$\vert e_{10}\rangle\equiv 0.$ 

Finally, a comparison of the matrix elements    
$\langle 0_1,0_3|{\rm Tr}_{2,E}\,[\vert\Gamma\rangle \langle\Gamma\vert]\, |0_1,1_3\rangle= 
\frac{1}{6}\, \langle e_{00}|e_{11}\rangle$ and 
 $\langle 0_1,0_3|{\rm Tr}_{2}\,[\vert\eta\rangle \langle\eta\vert]\, |0_1,1_3\rangle~=~ 
\frac{1}{6}$ 
lead to $\langle e_{00}|e_{11}\rangle=1$ or $\vert e_{11}\rangle\equiv \vert e_{00}\rangle.$ 
Thus,  the extended pure state (\ref{ws3}) should take the form 
$\vert \Gamma\rangle\equiv \vert \eta\rangle \, \vert e_{00}\rangle.$ 
In other words,  the three qubit pure state $\vert \eta\rangle$  is uniquely determined by its two-qubit reduced 
systems and is therefore, reducible. 

This illustrative example clearly projects out the contrasting   
irreducibility features of two SLOCC interconvertible states (\ref{GHZ}) and (\ref{Wsup}) of the {\em same}  
entanglement family $\{D_{1,1,1}\}.$ 

In conclusion, we have investigated the connection between SLOCC equivalence of pure states with the  
irreducibility of their correlations. With the help of representative examples of three 
qubit permutation symmetric states we have shown that interconvertibility does not necessarily imply 
irreducibility of 
correlations.


\begin{thebibliography}{0}
\bibitem{Majnote} Every term in the symmetrized superposition (\ref{Maj}) -- after effecting the identical 
rotation $R^{-1}_l\otimes R^{-1}_l\ldots \otimes R^{-1}_l$ -- contains at least one ``up" state
$\vert 0\rangle$, since $R^{-1}_l\vert \epsilon_l\rangle=\vert 0_l\rangle$  and so, the projection $\langle 
1_1,1_2,\ldots, 1_N \vert~ 
R^{-1}_l\otimes R^{-1}_l\otimes 
\ldots\vert \Psi_{\rm sym}\rangle$ of the rotated state in the ``all-down" direction
 vanishes.
\bibitem{WDicke}  An ILO $A\otimes A\otimes A$  with  
$A=\left(\begin{array}{cc} 1 & -\frac{c_0}{c_1} \\ 0 & 1 \end{array}\right)$,  
on the state $\vert D_{2,1}'\rangle$ transforms it to the three qubit W state.
\bibitem{noteAKR} Comparing (\ref{Wfin}) with (\ref{omega}) we readily find that 
$\vert e_{01}\rangle \equiv 0,$  and  $\vert e_{11}\rangle =\vert e_{00}\rangle$ leading to a simplified 
structure,  
$\vert\Omega\rangle=\vert {\cal D}'_{2,1}\rangle\, \vert e_{00}\rangle +
\vert \phi_1\rangle\vert 0_3\rangle\vert e_{10}\rangle$.
Now, requiring that $\rho_{1,2}={\rm Tr}_{3,E}\, \left[\vert\Omega\rangle\langle \Omega\vert\right]= 
{\rm Tr}_{3}\, \left[\vert {\cal D}'_{2,1}\rangle\langle {\cal D}'_{2,1}\vert\right],$ 
we readily find that $\vert e_{10}\rangle=0$ and $\langle e_{00}\vert e_{00}\rangle=1.$ This leads to 
 the desired result $\vert\Omega\rangle=
\vert {\cal D}'_{2,1}\rangle\, \vert e_{00}\rangle$ i.e., the reduced 2 party states  determine the pure state 
$\vert {\cal D}'_{2,1}\rangle$ 
uniquely.
\end{thebibliography}
\end{document}